# The Role of Spreadsheets in Clinical Decision Support: A Survey of the Medical Algorithms Company User Community

Simon Thorne sthorne@cardifmet.ac.uk
Cardiff Metropolitan University, the Medical Algorithms Company

**Abstract**

This paper presents and discusses the results of a small scoping survey of Clinical Decision Support System (CDSS) users from the Medical Algorithms Company website which hosts 24,000 different CDSS. These results are analysed, discussed, and compared with other similar studies and contribute to the wider understanding of how CDSS impact on clinical practice. The results show that CDSS provided by Medal are being used by clinical professionals in a variety of settings, both as an operational tool and as a research and reference tool. Whilst these tools are implemented and executed in a database, the initial logic is worked out on a spreadsheet. The paper describes that process and examines some of the results of the survey.

## 1.1 Introduction

This paper considers the Medical Algorithms Company user base and their approaches to making use of the Clinical Decision Support Systems (CDSS) and associated resources offered by MedAl. The paper presents and analyses results from a small survey (P=150) conducted with the users that reflects on the impact that such resources have on clinical practice.

## 1.2 The Medical Algorithms Company (Medal)

The Medical Algorithms Company is an organisation who specialise in converting algorithmic medical research, published in high quality peer reviewed medical journals, to small computer programs to assist in a wide variety of clinical decision making scenarios. The CDSS offered by Medal are CSS files executed in a database, the user can input patient and condition data, calculate an answer, and output a recommendation. Medal has approximately 24,000 different algorithms over a wide range of medical complaints. These range from Body Mass Index (BMI) through to how much anti-venom to administer in the event of a rattlesnake bite. Medal CDSS are accessed either through a web based interface or through an IOS/Android application. Medal facilitates users through a number of options, there are 4500 CDSS that can be accessed publicly either through the website or the cell application. Medal also offer Application Programming Interface (API) access to the full 24,000 CDSS which organisations can integrate into their own infrastructure.

## 1.3 Medal and spreadsheets

Iyengar [2009] discusses the now historical approach taken by Medal to provide CDSS to the clinical community via spreadsheet models. Medal at that time had 13,500 different CDSS spreadsheets across 45 medical specialties. These spreadsheets could be downloaded and used as needed. The rationale behind this is that spreadsheet technology is pervasive and can be used by practically anyone, hence this allows provision of important medical knowledge to many potential users. It was also thought that users of the spreadsheets could adapt and modify them to their specific needs which could facilitate an ever closer match of resource and need but also presents a potential source of error. Eventually it was judged that spreadsheet models are tooopen to error, piracy and misinterpretation and hence Medal considered other modes of deployment for CDSS.

As Medal has progressed as an organisation, the approach to providing CDSS has also evolved. Medal no longer provides spreadsheets for download or calculation, instead all CDSS are implemented and executed in a database with validation controls place on input and a better user interface than previous spreadsheet models. CDSS are available to users through a web interface, mobile applications and for large scale processing through API. However, spreadsheets are still at the heart of the development and provision of new CDSS by Medal, spreadsheets are used as a tool for developing the initial model and ensuring that the input output calculations reflect the clinical knowledge accurately [Shiverly and Harl, 2017]

**1.4 CDSS development process at Medal**

The development of CDSS at Medal is achieved through multiple identification, development and testing steps.

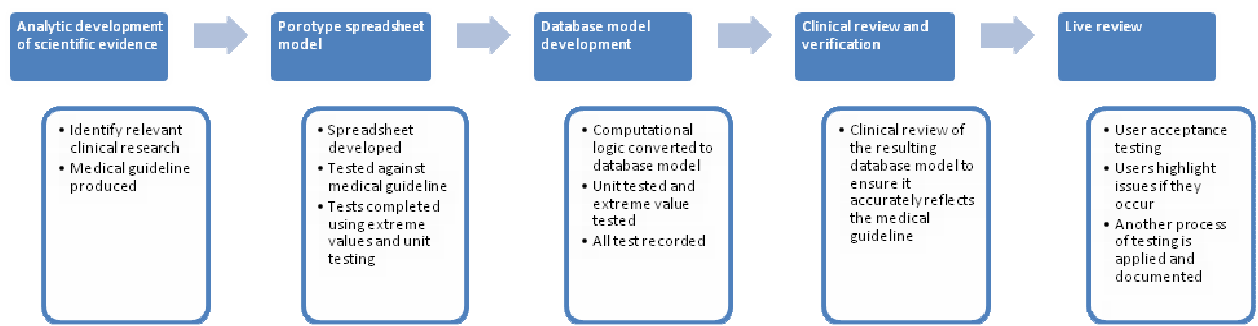

The first step of delivering a new CDSS is to identify suitable algorithms published in high quality peer reviewed medical research journals. This process requires a clinical expert to identify potentially interesting algorithms to implement. Once a paper has been identified, the knowledge contained in the paper is separated into 'chunks' as described by Simon [1974]. Chunks are a unit of knowledge measurement, Simon [1998] describes that most experts can accommodate between 50,000 to 100,000 chunks of information on a particular subject. Simon [1998] discusses the English language to contextualise chunks:

> "English speakers are experts on the English language — we have stored over 100,000 familiar chunks, which are called words. When we see them in a text, we recognize them and retrieve their meanings from memory"

In the case of algorithmic medical research, a single chunk is an equation that calculates part of the outcome. Between 5 and 20 chunks then form a "medical guideline" which is a document that aims to guide a clinician on decision making and best practice on the diagnosis, treatment, prognosis, risk/benefits and management of a particular condition. Medical guidelines are a key tool in clinical practice and are used extensively by clinical professionals across the world. The resulting medical guideline is abstracted in word, containing references and annotations all based on the style and conventions of Chemical Abstracts.

Once all of the chunks have been identified in the research and the medical guideline has been produced a prototype spreadsheet model of the guideline is created. At this point a clinical expert reviews the Excel model to ensure that it reflects the medical guideline created. The model is then tested by unit testing all input and output paths of the model and by testing extreme values and ranges. The test data and outcomes are stored for future reference. If these testsare passed then the spreadsheet is given to a programmer who implements the computational logic in a database. Once the database implementation is completed the database model is medically verified by the clinical team using the guideline and prototype spreadsheet to cross reference the database. If these tests fail, the process restarts at the medical guideline level.

If these tests pass, the algorithm can then become a deliverable module in the database and is documented. Once enough chunks have been identified and implemented, they are linked together in a modular fashion. Modular development allows reuse of code in algorithms and allows easy changing of components without the need to completely redevelop the entire solution. The clinical knowledge contained in modules is broken down into basic science, diagnosis, therapy, monitoring and patient concerns. Each of these has subtopics. The modules that are built can correspond to these. In theory, you can build a complete process addressing a specific clinical issue like diabetes, osteoarthritis and others. The final part of the process is to consider the language and notation issues from the medical guideline and then the CDSS is delivered.

It is thought that there might be a million chunks of knowledge currently in healthcare although several hundred CDSS would probably satisfy most CDSS users. The use of CDSS allows clinicians to be generalists who can through the use of CDSS knowledge to assess a wider range of clinical problems and therefore reduce the number of referrals needed. For instance a General Practitioner could treat nutrition problems with the right tools rather than referring a patient to another specialist. Therefore, there is significant cost savings that can be gained from using such resources. The process of developing CDSS in this fashion allows minimal costs to the organisation resulting in maximising the number of CDSS that can be provided to the clinical community.

### 1.5 The Medal user survey

A scoping survey was carried out by Medal with its users to gain more insight into the profile of the users, how the resources are used and what controls and checks are used by individuals and organisations.

### 1.6 Research questions

The principle aim of the survey was to answer the following research questions:

1. How are Medal CDSS and associated resources used by clinical professionals in terms of: Frequency of use; Experience and training levels with CDSS; Approaches to selection of CDSS; Validation and type of use; Importance of the resources to individuals and organisations and standards imposed on CDSS use by organisations.
2. What specific benefits and disadvantages do users perceive in utilising CDSS in clinical medicine and how do their assessments compare with our understandings from the literature?
3. How does the knowledge contained in CDSS impact on clinical practice and decision making?

However, for the purposes of this paper, not all research questions and data will be explored. This paper shall focus on the issues that are relevant to spreadsheet risk and approaches to managing such risk.

In order to better align this research with spreadsheet risk, the following additional research question is identified.

4. From the Medal user data on spreadsheet based CDSS use, what knowledge from spreadsheet risk management could benefit organisations and individuals making use of spreadsheet based CDSS?

### 2.0 Computing in Medicine

Computing in medicine has multiple applications, some that are well established such as the use of computers in medical imaging, and others that are still not fully exploited such as Electronic Health

Proceedings of the EuSpRIG 2017 Conference "Spreadsheet Risk Management" ISBN : 978-1-905404-54-4
Copyright © 2017, EuSpRIG European Spreadsheet Risks Interest Group (www.eusprig.org) & the Author(s)

Records (EHR) and CDSS. Computing power has the potential to revolutionise medicine in both efficiency and through communication of knowledge Kawamoto *et al.* [2010b]. Some computer systems are aimed at providing clinical support to medical professionals, some check potential drug-drug interactions, some provide reminders and alerts for clinical workflow and others are designed to manage patient data such as EHR.

## 2.1 Electronic Healthcare Records

Electronic Healthcare Records (EHRs) are a computer based replacement of the paper charts and records associated with patient care. EHRs offer more a more cohesive and centralised approach to gathering and managing patient data and in theory are more efficient and reliable than paper equivalents. EHRs have the potential to unite a number of disparate computing resources used in medicine, for instance pooling patient data, relevant drug and intervention reminders, clinical workflows and Clinical Decision Support Systems (CDSS) results all at the point of care. However, currently there is little integration of these systems [Kawamoto*et al.* 2010b] and the use of CDSS in particular remains ad-hoc as a result.

## 2.2 Clinical Decision Support Systems

Clinical Decision Support Systems (CDSS) are defined as software artefacts produced to assist in clinical decision making scenarios. Clinical decision making scenarios encompass a diverse range of activities that include but are not limited to: Workflow process clinical decision support; Research and reference for particular clinical problems; As an educational tool for learning and teaching medicine; As a tool for evaluating clinical problems by non clinical professionals; As a tool for patients. CDSS users include trained clinical professionals, training clinical professionals, non-clinical professionals in medical industries (such as Electronic Healthcare Records Companies and Medical Insurance) and patients.

## 2.3 Methodology

This section considers the design of the scoping survey in terms of questionnaire design, sampling, and delivery of the survey.

## 2.4 Questionnaire design

A questionnaire was chosen as the research instrument for this study since it is ideal at gathering shallow broad information ideal for a scoping survey. The survey serves two distinct purposes, firstly learn more about the Medal userbase and secondly to evaluate the impact of Medal resources on clinical practice.

The questionnaire will examine the user base along the following lines of enquiry: demographics, experience and frequency of use of the resources, approaches to selecting algorithms, validation and standards, importance of resources (organisational and individual), modes of access to resources, efficiency and specific benefits and limitations to individuals.

## 2.5 Participant recruitment and sampling

The sampling strategy for this survey is a clustered random approach. The cluster identified were users of the Medal website but no other filtering was applied, hence the random selection of the visitors or members of the site. This approach was chosen since the survey was aimed at discovering more about Medal users and, more broadly, CDSS users in general. Participants were recruited via a message on the splash page of the main Medal website and through an invitation to participate in the survey sent in the weekly email blog. Medal has around 40,000 registered users although it is not known how active those members are. The Medical Algorithms Company's website receives around 7000 unique hits per month but again little is known about those visitors other than they are unique.

Hence the potential population is relatively large, 150 responses were obtained over a period of around 4 months.

## 3.0 Results

### 3.1 Demographics

The demographics of the survey show that the large majority of the participants (40) were from the United States. The UK, India and Italy also featured 10 and 9 responses respectively. The remaining 91 responses were distributed amongst 33 other countries demonstrating the international nature of the resources Medal provide. 150 responses were gathered in total.

When asked about their occupation, 78% of the participants stated that their occupation was either Clinician, Healthcare Professional, Surgeon, Medical student or Nurse.

### 3.2 Experience, Training and frequency of access

This section considers user levels of experience with the CDSS tools, relevant training, the frequency of visits to the Medal site and how many calculators might be accessed in a typical single day.

Figure 6 shows the self perceived levels of experience with the Medal CDSS, as can be observed, over three quarters of participants indicated they had either little or experience with the resources or that they had beginner level experience. This finding is reflected in the literature [Kawamoto *et al.* 2010b], the provision and uptake of training is a major issue in the widespread adoption of CDSS. Whilst these resources are not spreadsheets as such, they are based on spreadsheet models.

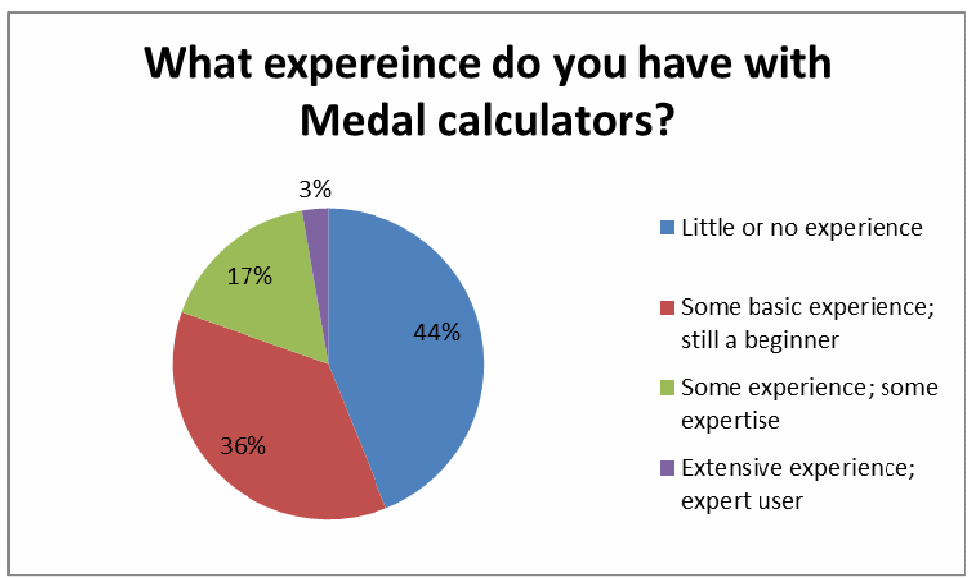

**Figure 1 Perceived experience with CDSS**

Figure 1 details the training received by participants in the use of CDSS, over three quarters of the respondents either have no training or are self-taught, hence most do not have professional training in use of CDSS. The literature reflects this point as appropriate training provision and uptake is poor amongst clinicians [Kawamoto *et al.* 2010b]. This also reflects the data in figure 2 which shows that most Medal users consider themselves as either having little or basic levels of experience with the resources.

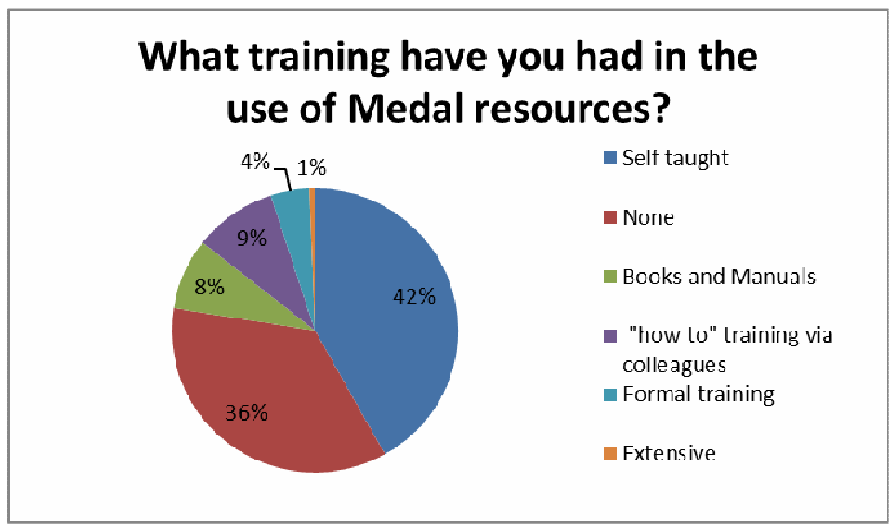

**Figure 2 Training in Medal resources**

When asked about frequency of access, 19% of participants access medal resources daily, 37% access the resources on a weekly basis and 20% access once a month. This is interesting since it shows that about 56% of the participants are accessing the Medal resources at least once a week. This shows that the resources are used frequently by the participants which would suggest that the CDSS resources are valuable to the participants either in their own personal work or through their organisations aims.

When asked about how many Medal calculators the participants accessed in a typical day, 30% said that in a typical day there would not access Medal resources. However, 58% of the participants use at least one calculator a day and 15% access 4 or more resources per day. This suggests that the Medal resources are valuable to the participants and that the penetration of CDSS in various medical settings is becoming more common. In total, 46% of the participants indicated that they worked in a hospital setting, of that46% only 15% said that they didn't use the resources, leaving 31% as 'active users'. Of the 31% active users, 41% said they access at least one Medal calculator a day, 42% accessed 2 a day, 7% accessed 3 a day and the other 10% said they accessed 4 or more a day.

These findings are interesting since it suggests that the Medal CDSS are being used operationally in hospitals to assist in diagnosis or research of conditions. Kawamoto [2005] discusses the lack of inclusion of CDSS in clinical workflow processes as a major barrier to the uptake of such resources, some of the Medal users are getting around this by accessing Medal calculators on mobile devices and desktop computers. Of the 37% who use the resources at least once a day, 67% say they use an Android or IOS cell/Mobile/Smart phone for access, 19% use desktop computers for access, 8% use laptops and 5% use Android or IOS tablets. This data seems to confirm that hospital workers are using mobile devices to access these resources operationally to assist in their day to day duties.

**3.3 Validation of CDSS results and organisational standards**

Validating findings obtained from CDSS is an important step in the use of CDSS to enhance medical care. Equally, the organisations attitude towards the use of CDSS is another important part of the safe and efficient application of CDSS technology. This coming section explores those issues.

Participants were asked about approach used to validate results obtained from Medal CDSS. The most common approach to validating results is to check CDSS data against published relevant medical literature as answered by 44% of the participants; 28% answered they accept the calculators were correct; 18% said they cross check the results with colleagues and 6% sought the opinion of a superior. This means that of the 150 total participants, 69% employ some form of check to validate the results. The validation process is made easier if the CDSS are part of the clinical workflow and hence standards can be applied to ensure validation and evaluation of results (Kawamoto *et al.* 2010b].

Two questions asked the participants about the importance of the resources offered by Medal both to the organisation they work for and to themselves as individuals. When asked about the importance to the organisation, the most frequent response was "moderately important" with 40% ofparticipants indicating so; 17% said the resources were "very important"; 3% said "critical" and 26% said the resources were "unimportant". It would seem that organisations do rely on these resources to some extent 60% of participants said that Medal resources were at least moderately important to the organisation.

When asked about the importance of the resources to individuals, the most frequent response was "moderately important" with 49% of participants indicating so; 22% said "very important"; 15% said "unimportant" and 2% said "critical". In comparison to responses of organisational importance, the results are very similar but fewer said that the resources were unimportant. Perhaps one explanation for this is that the resources have value to individuals as tools to support their activities but organisations have not yet widely recognised the usefulness and importance of the resources fully. A common problem reported in the literature is the lack of CDSS as part of the workflow process and perhaps these results are indicative of this point [Kawamoto*et al.* 2005, Kawamoto*et al.* 2010a, Eichner and Das 2010, Castaneda *et al.* 2015, Bakken *et al*, 2007].

When asked about organisational standards that govern the use of Medal resources, see figure 3, 66% said there were no standards set; 16% said there were informal unwritten guidelines, 14% said they have basic written standards and 5% said there were detailed written standards. Consider these results in light of figures 12 and 13 which both show that Medal resources are important to both organisations and individuals. It would seem that organisations have failed to acknowledge and incorporate Medal resources, and other non Medal CDSS, into their workflows and in turn, internal standards for use. The development and deployment of standards associated with CDSS are important since ad-hoc use of such systems lends itself to misunderstandings and sub optimal efficiency. This finding is echoed by Kawamoto *et al.* [2010b] who discuss the broad lack of standards associated with CDSS use and the lack of formal uptake of CDSS systems in the medical community. In addition, this data suggests that CDSS use is unknown to clinical organisations which is similar in nature to how spreadsheets tend to make up a large part of businesses computing infrastructure without explicit knowledge of the organisation. Panko [2013] described spreadsheets as the dark matter of corporate IT infrastructure, perhaps CDSS will become the dark matter of clinical computing.

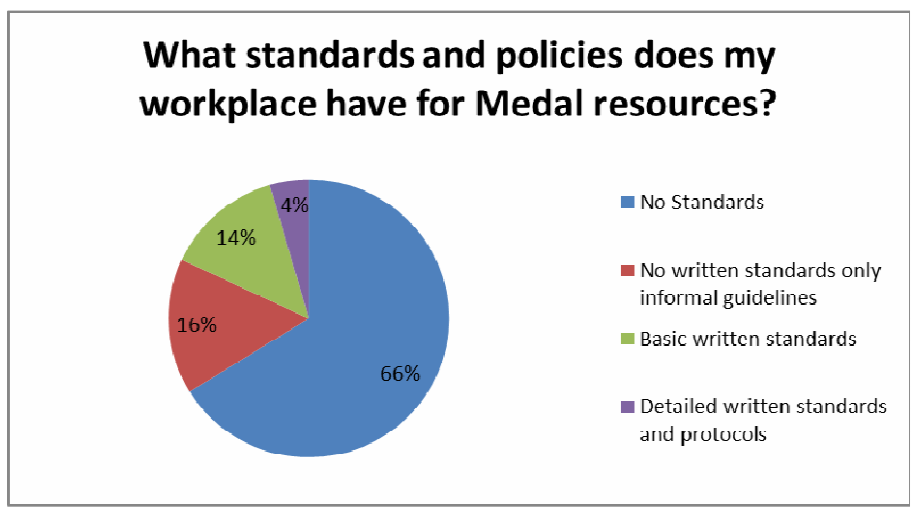

Figure 3 Standards and Policies that govern CDSS use

When the participants were asked about observation of standards at their organisation, most participants indicated that they do observe standards set, see figure 4. This suggests that if organisations had suitable policies, they would be observed. Hence organisations seem to be trailing behind the medical community who are making use of such resources.

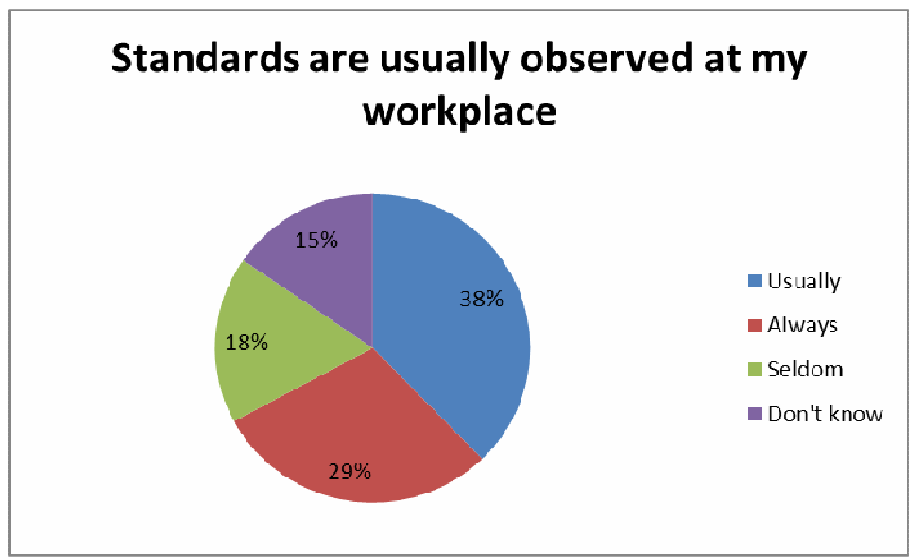

**Figure 4 Observation of standards and policies**

**3.4 Benefits and Limitations to Medal resources**

Figure 5 shows how the participants use the Medal Calculators, 36% indicated they use the calculators as an educational learning and teaching tool; 34% indicated they use the calculators as a research and reference tool; 30% indicated that they use the calculators as a day to day decision support tool and <1% indicated research and development. These results show that Medal calculators are used in two basic modes, either as a research, reference, educational or teaching tool, or they are used operationally to make day-to-day decisions.

Further, if one isolates the 61% of participants who indicated they worked in a hospital or medical practice , 66% say the use as research and reference tool; 63% say they use the calculators as a day to day decision support tool and 73% say they use the calculators as an educational learning or teaching tool. This suggests that those who work in hospitals or medical practices are using multiple strategies to exploit the resources since several indicated that they use Medal CDSS in all categories of the question posed. Further, since these participants work in either a hospital or general medical practice, one can infer that staff are operationally using Medal CDSS to assist in the diagnosis of conditions, as research and reference material and as an educational learning or teaching tool at the point of care. This provides further evidence that CDSS have a practical and essential operational use.

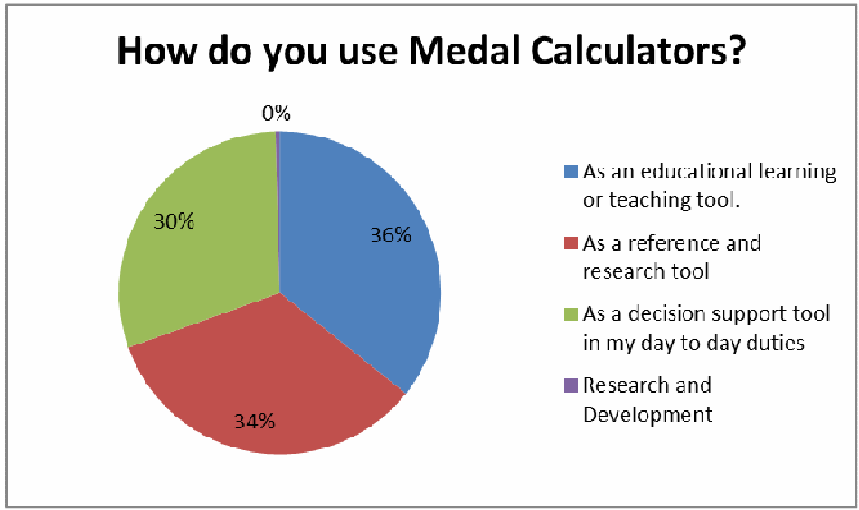

**Figure 5 Approaches to use of Medal CDSS**

When participants were asked how Medal calculators impact on work performance: 36% said they can broaden and deepen their knowledge; 33% said they achieve greater time efficiency; 17% said the resources allow them to communicate with patients more efficiently; 12% said the resources allow them to communicate medical conditions to clinical and nursing students, 1% said validation of diagnosis and 1% said more efficient research. See figure 6. Clearly, the Medal calculators offer multiple benefits to the participants as both an operational front line tool but also as a study and reference tool to increase an individual's knowledge and to teach others.

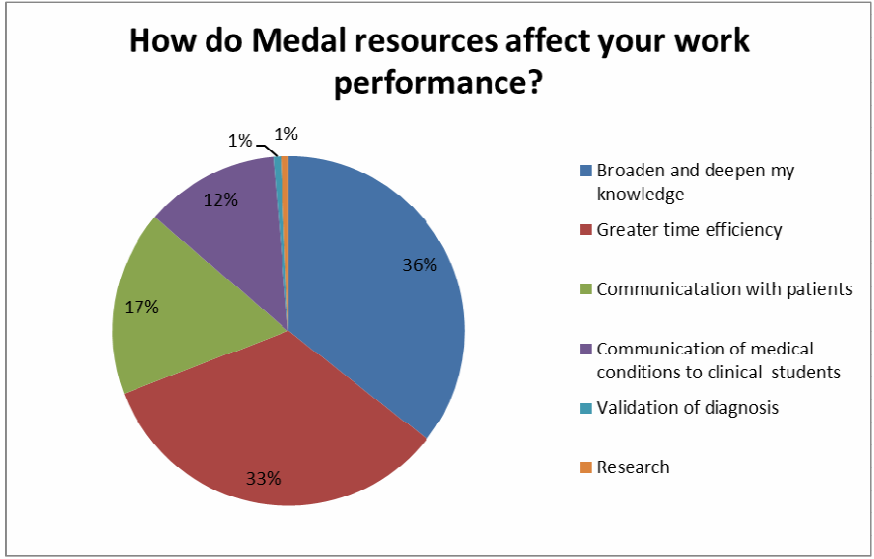

**Figure 6 Effect of Medal CDSS on performance**

## 3.5 Discussion

This next section considers the results of the survey and explores how these issues relate to spreadsheet risk management, both in terms of risk presented using these resources but also what can be bi-directionally learnt from CDSS use and spreadsheet risk management.

Although this survey relates to CDSS provided by Medal, it is reasonable to assume that Medal are not the only source of CDSS used by participants and that spreadsheets are probably used in addition to these resources across the clinical community. Croll and Butler [2006] explored clinical spreadsheets publicly available on the internet and found over 800 references to spreadsheet applications by searching the pubmed database (https://www.ncbi.nlm.nih.gov/pubmed/). In addition, although the questions posed relate directly to the Medal CDSS resources, some questions have wider implications such as the development and adoption of organisational standards for CDSS resources.

## 3.6 Risks and mitigation of CDSS use

From the data it is clear that there are certain risks that can be identified in the practice of individuals and organisations relating to CDSS. There are also some corrective and preventative actions that can be taken to mitigate such risks which will be discussed in turn.

Figure 1 shows that over three quarters of the participants consider themselves to have either little or no experience or beginner level experience. In addition, when asked about training over three quarters said they either have had no specific training for such resources or that they have self-taught themselves to use the resources, see figure 2. This highlights a deficiency in current education practice since such resources are likely to become increasingly important in future medical care. There is also an interesting parallel to be drawn with spreadsheet training. Research in spreadsheet development shows that most users either have no formal training or they have self-taught, Lawson *et al.* [2006] discuss a survey of 1597 MBA alumni from various HE institutions in the United States and Europe. When asked about training, 40% said that they had no formal training in the use of spreadsheets and that being 'self-taught' was the most common response. This issue is echoed in other research too such as [Taylor, 1998] which shows that most 'end users' do not have formal training in the design and implementation of spreadsheet tools. This should be rectified by organisations devising and offering clinical professionals training in the use and interpretation of CDSS.

When asked about validation of results obtained from CDSS, 70% employ some form of double checking either through checking the medical literature, seeking the opinion of colleagues or the opinion of a superior. However, 30% said that they accept that the calculators were correct implying that they employ no other form of checking. Whilst medical professionals are highly trained, this presents a risky behaviour amongst the participants since misconceptions, misinterpretations and overconfidence can affect even highly trained and experienced professionals [Lusted 1977, Oskamp 1965] which could lead to poor decision making with significant consequences. Overconfidence is a difficult issue to mitigate since it is so widespread, Panko [2003] discusses approaches to reduce overconfidence in spreadsheet development and finds that informing participants that the spreadsheets produced could contain errors seemed to reduce overconfidence and improve accuracy. In CDSS the same thinking could be applied with the application of checks through a wider set of standards and controls that links CDSS use with EHR.

When the participants were asked about the importance of CDSS resources to both the organisation they work for and to their own personal role in the organisation, participants indicated that CDSS results were moderately important to both their organisation and personal activity. When asked how many CDSS resources they make use of in a typical week, 75% said they use at least one CDSS. When asked about how the resources are used, see Figure 5, 42% said they use these resources to assist in day to day decision making duties. It is therefore reasonable to assume that Medal CDSS are important to organisations and that in the future the importance of these resources is likely to increase. Although CDSS resources have been available for some time, the use of CDSS in medicine is relatively new phenomena. In time it is likely that such resources will become critical to organisations in the same way the spreadsheet applications are now deemed critical to business decision making. This implies that CDSS use will increase significantly as the technology becomes more common in clinical medicine as will the number of issues regarding use of the technology. Hence clinical organisations should pay special attention to setting standards for CDSS use and formalise the current ad-hoc approach to CDSS deployment, use and integration.

Figure 3 shows that 66% of organisations that participants work for do not have any standards that govern the use of CDSS technology. In combination with the lack of training in such resources, this presents a risk that CDSS could be misused or misinterpreted. This could result in inefficiencies in decision making and potentially more severe consequences through misapplication of knowledge. Clearly this is an organisational management issue that needs to be addressed at the highest levels of the organisations in question to avoid any unwanted consequences of misuse of the technology [Kawamoto *et al.* 2010b]. Hand in hand with the lack of standards is the current lack of integration of CDSS with other clinical systems such as EHRs, this issue is commonly reported in CDSS literature [Hunt *et al.* 1998, Kawamoto *et al.* 2010b, Das and Eichner 2010, Ash *et al.* 2003, Garg *et al.* 2005]. Integration is critical since it streamlines the use of CDSS resources in clinical settings and allows organisations to monitor and control the use of CDSS. This first step is critical in both fully exploiting the potential of the resources but also ensuring that the resources are used in a safe and appropriate manner.

For the full potential of CDSS to be realised, organisations must concentrate on integration of CDSS with other clinical systems and develop suitable policies on adoption and use [Moja *et al.* 2014, Thomas and Coleman 2012]. A fully integrated system could facilitate Evidence Based Medicine (EBM) on a patient by patient basis, providing a dossier of evidence for a decision reached. Through the use of Medal's API, full integration of CDSS, EHR and EBM is possible however it seems that organisations are not adopting this integration currently. In the pharmaceutical industry, the Food and Drug Administration (FDA) impose strict controls on electronic artefacts used in the analysis of drug trials via title 21 CFR Part 11 [FDA, 2017].

Title 21 CFR Part 11 protects against security violations and ensures the reliability of electronic records. The legislation demands companies provide evidence of: audits, validation, electronic signatures and documentation for any software artefact. The legislation also dictates that electronic artefacts be stored in a secure server so that once the artefact has been created and audited, it cannot be changed without authorisation and access is limited by 'lock and key'. Authorisation would be needed to implement any changes with subsequent auditing, validation and documentation. Such a system could be adapted for the integration of CDSS as part of EBM. For instance, a CDSS could be used to assess a patient's condition on a particular medical issue, once the calculation is complete, a complete copy of all input and output data would be kept on a secure server. This would greatly improve reliability and accountability and provide an audit trail for the decision reached. Indeed, it is recommended that spreadsheet modelling follows the same approach to reduce fraud and increase accountability [Thorne,

## 3.7 Conclusions

This paper has described the process of CDSS development at Medal which is achieved via a prototype spreadsheet. Once the spreadsheet has passed a number of tests, the logic is then implemented into a database and is executed via a web interface, mobile application or API access.

The paper then considers the users of Medal CDSS and their attitudes and experiences on a number of important issues. Interestingly, there are many similarities with the spreadsheet modelling world. For instance: training in the use of specific resources is scarce; most organisations do not seem to be aware of extensive CDSS use in their organisations; the level of importance attached to the resources by the users is at least moderately important and organisations do not have any policies that govern their use. Although this survey does not cover individuals creating CDSS, only using, this set of conditions is rather similar to spreadsheet practices evident in organisations today. They are uncontrolled, ad-hoc, ignored by the management of the organisation and increasingly critical to users.

## References

Ash, J. S., Sittig, D. F., Campbell, E. M., Guappone, K. P., & Dykstra, R. H. [2007]. Some Unintended Consequences of Clinical Decision Support Systems. *AMIA Annual Symposium Proceedings*, *2007*, 26–30.

Castaneda. C, Nalley. K, Mannion. C, Bhattacharyya. P, Blake. P, Pecora. A, Goy. A, Stephen.S, [2015], Clinical decision support systems for improving diagnostic accuracy and achieving precision medicine, *Journal of Clinical Bioinformatics*, 5:4, doi 10.1186/s13336-015-0019-3

Croll G. J, Butler R. J, [2006] Spreadsheets in clinical medicine: A public health warning. *Proceedings of the European. Spreadsheet Risks Interest Group, Cambridge, UK*.http://www.eusprig.org/2006/spreadsheets-in-clinical-medicine-warning.pdf

Das M, Eichner J., [2010] Challenges and Barriers to Clinical Decision Support [CDS] Design and Implementation Experienced in the Agency for Healthcare Research and Quality CDS Demonstrations (Prepared for the AHRQ National Resource Center for Health Information Technology under Contract No. 290-04-0016.) AHRQ Publication No. 10-0064-EF. Rockville, MD: Agency for Healthcare Research and Quality.

Garg. A, Adhikari. N, McDonald. H, Rosas-Arellano. M, Devereaux. P, Beyene. J, Sam. J, Haynes. B, [2005], Effects of Computerized Clinical Decision Support Systems on Practitioner Performance and Patient Outcomes A Systematic Review,*JAMA,* 293(10):1223-1238.

FDA, [2011], CFR - Code of Federal Regulations Title 21 Part 11 2011.https://www.fda.gov/downloads/regulatoryinformation/guidances/ucm125125.pdf

Hunt D, Haynes R, Hanna S, Smith K., [1998], Effects of Computer-Based Clinical Decision Support Systems on Physician Performance and Patient Outcomes A Systematic Review,*JAMA,* 280(15):1339-1346

Iyengar. S, [2009], The Medical Algorithms Project, *Proceedings of the European. Spreadsheet Risks Interest Group*, EuSpRIG, Paris, France.https://arxiv.org/ftp/arxiv/papers/0908/0908.0932.pdf

Kawamoto, K., Del Fiol, G., Lobach, D. F., & Jenders, R. A. [2010]. Standards for Scalable Clinical Decision Support: Need, Current and Emerging Standards, Gaps, and Proposal for Progress. *The Open Medical Informatics Journal*, *4*, 235–244. http://doi.org/10.2174/1874431101004010235

Kawamoto, K., Del Fiol, G., Orton, C., & Lobach, D. F. [2010]. System-Agnostic Clinical Decision Support Services: Benefits and Challenges for Scalable Decision Support. *The Open Medical Informatics Journal*, *4*, 245–254. http://doi.org/10.2174/1874431101004010245

Kawamoto, K., Houlihan, C. A., Balas, E. A., & Lobach, D. F. [2005]. Improving clinical practice using clinical decision support systems: a systematic review of trials to identify features critical to success. *BMJ*5 *: British Medical Journal*, *330*(7494), 765. http://doi.org/10.1136/bmj.38398.500764.8F

Lawson. B, Baker. K, Lynn. F and Powell. S, [2006], "A Survey of MBA Spreadsheet Users." Spreadsheet Engineering Research Project. Tuck School of Business. 2006. ‹http://faculty.tuck.dartmouth.edu/images/uploads/faculty/serp/survey_paper.pdf›.

Lusted. L, [1977], 'A study of the efficiency of diagnostic radiological procedures: Final report on diagnostics efficiency', Chicago: Efficacy Study Committee of the American College of Radiology

Moja, L., Kwag, K. H., Lytras, T., Bertizzolo, L., Brandt, L., Pecoraro, V. Bonovas, S. [2014]. Effectiveness of Computerized Decision Support Systems Linked to Electronic Health Records: A Systematic Review and Meta-Analysis. *American Journal of Public Health*, *104*(12), e12–e22. http://doi.org/10.2105/AJPH.2014.302164

Oskamp, [1965], 'overconfidence in case-study judgements', *The journal of consulting psychology*, 29, pp 261- 265

Shiverly. J and Harl. J, [2017], CDSS development processes at Medal, *personal communications with the author*

Simon. H, [1974], How big is a chunk?, *Science*, American Association for the Advancement of Science

Simon. H, [1998], What we know about learning, *Journal of Education Engineering*, 87, 343-348

Thomas. S and Coleman. J [2012], The impact of computerised physician order entry with integrated clinical decision support on pharmacist–physician communication in the hospital setting: a systematic review of the literature, *European Journal of Hospital Pharmacy: Science and Practice,* **19,** 349-354.

Thorne. S, [2013], The misuse of spreadsheets in the nuclear fuel industry: The falsification of safety critical data using spreadsheets at British Nuclear Fuels Limited (BNFL), *Journal of Organizational and End User Computing*, 25 (3), 20-31

Veseley, R. [2017] How Predictive Analytics Can Help Prevent Infection, Hhnmag.com. Available at: https://goo.gl/kgMgQi (Accessed: 11 May 2017).

Wanderer, J. P., Sandberg, W. S., & Ehrenfeld, J. M. [2011]. Real-Time Alerts and Reminders Using Information Systems, *Anaesthesiology Clinics*, *29*(3), 389–396. http://doi.org/10.1016/j.anclin.2011.05.003